\documentclass[10pt,aps,prl,showpacs,floatfix,floats,superscriptaddress,twocolumn]{revtex4-1}
\usepackage{graphicx,latexsym}
\usepackage{dcolumn}
\usepackage{amssymb,amsmath,bm}

\usepackage{amsfonts}
\usepackage{amsmath}
\usepackage{epstopdf}

\usepackage{epsfig}
\usepackage[dvips]{color}
\usepackage{hyperref}
\hypersetup{
    pdfnewwindow=true,       
    colorlinks=true,         
    linkcolor=blue,          
    citecolor=blue,          
    filecolor=magenta,       
    urlcolor=black           
}

\def\eq#1{Eq.\ (\ref{#1})}

\def\fig#1{Fig.\ \ref{#1}}

\begin{document}
\title{Transport signatures of top-gate bound states with strong Rashba-Zeeman effect}
 \author{Chi-Shung Tang}
 \email{cstang@nuu.edu.tw}
 \affiliation{Department of Mechanical Engineering, National United University,
 Lienda 2, Miaoli 36063, Taiwan}
 \author{Yun-Hsuan Yu}
 \affiliation{Department of Mechanical Engineering, National United University,
 Lienda 2, Miaoli 36063, Taiwan}
 \author{Nzar Rauf Abdullah}
 \affiliation{Physics Department, College of Science,
University of Sulaimani, Kurdistan Region, Iraq}
 \author{Vidar Gudmundsson}
 \affiliation{Science Institute, University of Iceland,
        Dunhaga 3, IS-107 Reykjavik, Iceland}



\begin{abstract}
We suggest a single-mode spin injection scheme in non-ferromagnetic
quantum channels utilizing perpendicular strong Rashba spin-orbit
and Zeeman fields.  By applying a positive top-gate potential in
order to inject electrons from the spin-orbit gap to the low-energy
regime, we observe coherent destruction of transport signatures of a
hole-like quasi-bound state, an electron-like quasi-bound state, or
a hole-like bound state features that are sensitive to the selection
of the top-gate length along the transport direction.
\end{abstract}


\maketitle

\paragraph{Introduction.}

Quasi-one-dimensional narrow constrictions in two-dimensional electron
gases (2DEGs) are among the most widely studied elements illustrating the
wave and spin nature of electrons for both application and fundamental
arenas for many
years~\cite{Wharam1988,vanWees1988,Thomas1996,Bardarson04}. Spintronics
utilizing the spin degree of freedom of conduction electrons is one of the
most promising paradigms for the development of novel devices for
applications from logic to storage devices with low power consumption.
For the purposes of spintronics applications, narrow band gap
semiconducting materials being integrated for verification of spin
transport would be of great
interest~\cite{Datta90,Governale02,Aharony2008}. This could be achieved
either through detection of the spin current or spin accumulation at the
edges of the device.  Generation, detection, and manipulation of electron
spin in mesoscopic systems has thus been the aim of spintronics for both
application and fundamental
arenas~\cite{Wolf2001,Aws2002,Malshukov03,Zutic2004,Malshukov05}.

For a built-in electric field perpendicular to the asymmetric 2DEG plane,
the momentum-dependent spin-orbit magnetic field is aligned perpendicular
to the quantum channel and in the plane of 2DEG (Rashba
effect)~\cite{Rashba60,Bychkov1986,Galitski13}.  The spin-orbit
interaction (SOI) has attracted much attention on account of its possible
applications in spin-based electronics since the Datta-Das spin transistor
was proposed~\cite{Datta90}. The essential requirement for spintronic
devices is to manipulate the free electron spins that can be achieved via
an external active control. Since Nitta {\it et al.}~\cite{Nitta1997}
showed that the Rashba SOI can be controlled, interest in utilizing the
Rashba SOI to manipulate electron spins in constricted systems has been
growing. The control of Rashba interaction was demonstrated to be material
and structure sensitive~\cite{Eldridge2008,Sheremet2016}.  The combination
of strong Rashba and weak Zeeman fields would induce a spin-orbit
gap~\cite{Pershin2004,Quay2010,Sadreev2013,Loss2014} or a helical
gap~\cite{Cayao15}. Very recent experimental and theoretical
studies~\cite{Manchon15,Kammermeier16,Heedt17} have successfully shown
that the presence of strong Rashba coupling in InAs nanowires can be used
for spin manipulation. In our previous
works~\cite{Tang2012,Tang2015,Tang2017}, we have considered delta profile
finger-gate controlled spin-polarized transport and thus did not need to
consider evanescent and propagating modes in the gate region.

\begin{figure}[t]
\includegraphics[width=0.44\textwidth,angle=0]{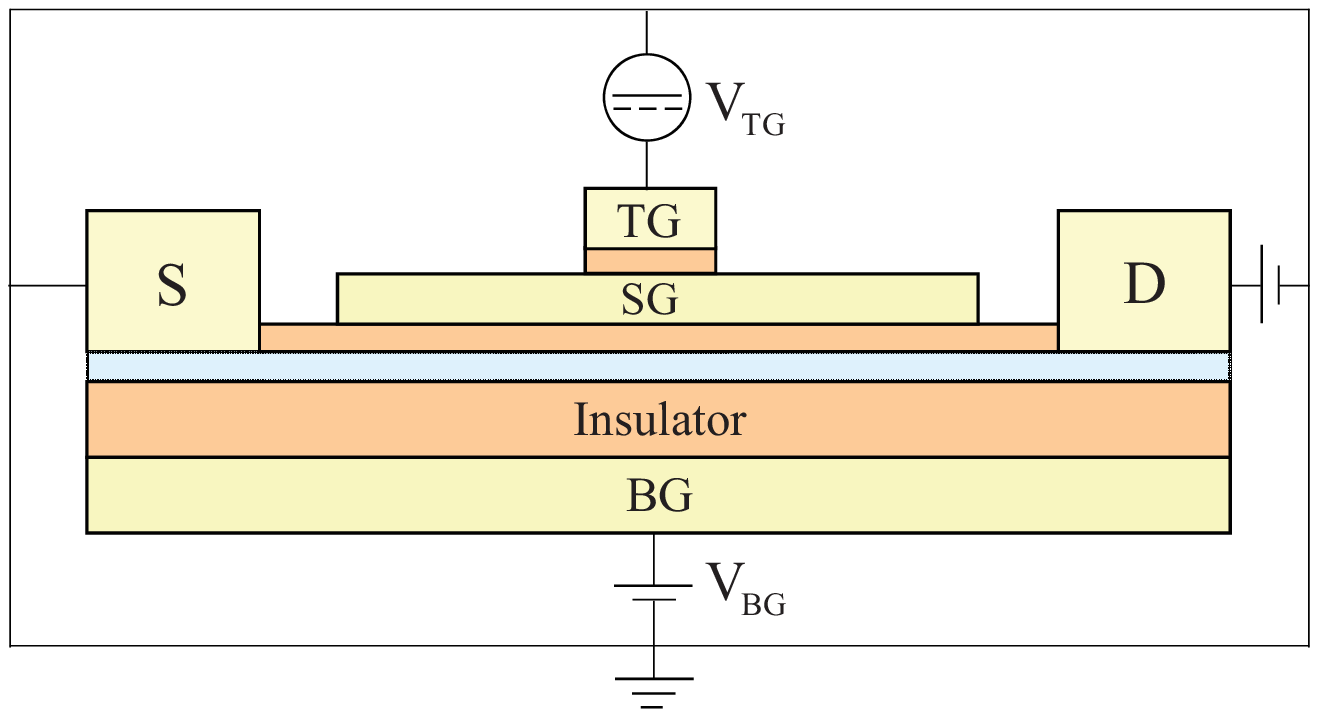}
 \caption{
Side-view of the considered system. A dc top-gate (TG) voltage
$V_{\rm TG}$ is applied to control the split-gate (SG) confined
quantum channel under a small dc bias between source (S) and drain
(D) electrodes. An additional back-gate (BG) voltage $V_\textrm{BG}$
is applied to tune the energy of the conduction electron. These
gates and electrodes are separated by insulators (orange).}
\label{fig1}
\end{figure}

In this Letter we consider a top-gate controlled quantum device (see
\fig{fig1}) under an in-plane oriented magnetic field parallel to the
direction of current flow.  A back-gate voltage $V_\textrm{BG}$ is applied
for tuning the Fermi energy of conduction electrons in the anisotropic
2DEG between the split-gates and the back-gate.  A similar experimental
set-up has been proposed~\cite{Kammermeier16}, in which the top-gate acts
upon a hexagonal cross-section wire forming a quantum point contact.

We shall show that around the spin subband minima or maxima,
Fabry-P\'{e}rot interferences between propagating modes would be
suppressed, and new transmission resonances with different
properties occur due to the coupling to spin-resolved evanescent
modes. The application of voltage at the top-gate results in
potential barrier in the channel and determines the shift of energy
dispersion in the gate region. As a result, within a single-mode
spin injection scheme, we observe that the spin-polarized dc
conductance spectra may reveal significant electron-like and
hole-like quasi-bound states (QBSs) as well as hole-like bound
states (BSs) features depending on the selection of the length and
the applied voltage of the top-gate.

\paragraph{Spin dependent transport.}

We consider a Rashba quantum channel in the plane of the 2DEG confinement
to be located at $z$ = $0$, and a magnetic field contributes the
Hamiltonian via the spin Zeeman term $\bf{H}_{\mathrm{Z}}$.  In the
absence of a top-gate, the system can be described by an unperturbed
Hamiltonian
\begin{equation}
 \mathbf{H}^0\left(x,y\right)
 = {\bf{H}}_{{\rm{1D}}}^{\rm{0}}\left( x \right)
 + {\bf{H}}_{{\rm{SG}}}^0\left( y \right)
 + {{\bf{H}}_{\rm{R}}} + {{\bf{H}}_{\rm{Z}}}\, .
 \label{H0}
\end{equation}
The first two terms describe an ideal quantum
channel~\cite{Tang2012,Tang2015,Tang2017}. In addition, the third term is
the Rashba effect induced by a built-in electric field due to asymmetric
2DEG confinement, and the last term denotes the Zeeman effect induced by
an in-plane magnetic field. Below, we employ the Fermi level as an energy
unit $E^*$ = $E_\textrm{F}$ and the inverse wave number as a length unit
$l^* = k_{\rm F}^{-1}$. Correspondingly, we define the Rashba coefficient
unit $\alpha^* = E^*l^*$ and magnetic field unit $B^*$ =
$E_\textrm{F}/\mu_\mathrm{B}$ for convenience. The Rashba interaction can
be expressed as $\mathbf{H}_{\rm R} = -2\alpha \sigma_y k_x$ in a narrow
constriction~\cite{Tang2012}. The Zeeman term $\mathbf{H}_{\rm Z}$ = $gB
\sigma_x$ is characterized by the half-Land\'e factor $g$ = $g_s/2$.

In this work we show how the spin-polarized electron transport in a
quantum channel is influenced by a dc top-gate (see \fig{fig1}).  The
top-gate electric potential energy is modeled by a rectangular form
$U_{\rm TG}(x)$ = $U_0 \theta(x)\theta(L-x)$ for simplicity where $U_0$ =
$-eV_{\rm TG}$ and $\theta$ is the Heaviside step function.  Plane wave
solution for the eigenvalues of the Schr\"{o}dinger equation gives the
spin dependent energy dispersion of the top-gate device
\begin{equation}
{E_n^\sigma = {\varepsilon _{n}} + k_x^2 + \sigma \sqrt {{{(gB)}^2}
+ {{(2\alpha {k_x})}^2}}} + U_{\rm TG}(x)
 \label{E-k}
\end{equation}
with $\sigma$ = $\{+,-\}$ denoting the spin-up ($+$) and spin-down
($-$) branches, and eigenvectors
\begin{equation}
{\left| {k_x^ \pm } \right\rangle} = \frac{1}{{\sqrt 2 }}
 \left( {\begin{array}{*{20}{c}}
1\\
{ \pm {e^{ - i\phi \left( {{k_x}} \right)}}}
\end{array}} \right)
 \label{eigenvector}
\end{equation}
with $\phi \left( {{k_x}} \right)$ = ${\tan ^{ - 1}}\left( {2\alpha
{k_x}/gB} \right)$.

In the absence of a magnetic field, the two lateral spin-split
subbands are lowered by a spin-orbit energy and shifted in
momentum-space by the spin-orbit wave number $k_\textrm{so}^{\pm}$ =
$\mp \alpha$ ($B$ = 0 T)~\cite{Heedt17}. For a given subband $n$,
the spin-orbit energy $E_{\rm so}$ = $\alpha^2$ can be defined by
the energy difference between the degenerate energy crossing point
$E_n^{\rm cross}$ = $\varepsilon_n$ at wave number $k$ = $0$ and the
subband bottom of the lower spin branch $E_{\rm bottom}^-$ =
$\varepsilon_n -\alpha^2$.

In the presence of a magnetic field, the two vertical spin-split
subbands are lowered by a $gB$-dependent spin-orbit energy and
shifted in momentum-space by the magneto-spin-orbit wave number
\begin{equation}
k_\textrm{so}^{\pm}(gB)=\mp \sqrt {{\alpha ^2} - {{\left(
{\frac{{gB}}{{2\alpha }}} \right)}^2}}\, .
\end{equation}
In the presence of the Zeeman term, the spin-orbit energy is
modified and can be defined by the energy difference between the
lower spin branch top $E_{\rm top}^{-}(gB)$ = $\varepsilon_n$ $-$
$gB$ at wave number $k_x = 0$ and the lower spin branch bottom
$E_{\rm bottom}^-(gB)$ = ${\varepsilon _n}$ $-$ ${\alpha ^2}$ $-$
$\left( gB/2\alpha\right)^2$. This gives a magneto-spin-orbit energy
\begin{equation}
E_\mathrm{mso} = {\alpha ^2} - gB + {\left( {\frac{{gB}}{{2\alpha }}}
\right)^2}\, ,
\end{equation}
in which both a linear and a quadratic term of the Zeeman factor
$gB$ appear.

We would like to bring reader's attention to the \eq{E-k} that only shows
the energy dispersions of the propagating modes. In order to calculate
both the propagating and evanescent modes associated with the multiple
scattering by the dc-biased top-gate, one has to consider the incident
electron with a given energy $E_n$ = $E - \varepsilon_n$ in the subband
$n$ by solving the quantum dynamic equation~\cite{Tang2015}.
\begin{equation}
k_x^4 - 2\left( E_n + 2\alpha^2 \right) k_x^2  + E_n^2 - (gB)^2 =
0\, .
 \label{kx_En}
\end{equation}
Solving this equation for a given energy $E_n$ allows us to
determine all wave numbers $k_x$, real or complex, corresponding to
propagating or evanescent modes, respectively.

In the quantum channel the top-gate boundaries couple spin-split
propagating modes to spin-flip non-propagating modes. Therefore, the
transport current causes localized modes to build up around the
top-gate edges.  A $4\times4$ top-gate propagation matrix method is
used to take four spin-split right-moving and left-moving states
into account, given by
\begin{equation}
\mathbf{P}_{\textrm{top-gate}} = \prod _{j = 1}^2
\mathbf{P}_{j,\mathrm{step}}\mathbf{P}_{j,\mathrm{free}}
 =
 {
 \left[ {\begin{array}{*{20}{c}}
 {{{\mathbf{p}}_{11}}}&{{{\mathbf{p}}_{12}}}\\
 {{{\mathbf{p}}_{21}}}&{{{\mathbf{p}}_{22}}}
 \end{array}} \right]
 }\, .
\end{equation}
Here two step propagation matrices (step-up and step-down) as well
as two free propagation matrices (with and without $U_0$) are
involved to incorporate multiple scattering at the two ends of the
top-gate. The transmission and reflection matrices are,
respectively, ${\mathbf{t}}$ = ${\mathbf{p}}_{11}^{-1}$ and
${\mathbf{r}}$ = ${{\mathbf{p}}_{21}}{\mathbf{p}}_{11}^{-1}$
involving spin-preserve and spin-flip contributions.  Taking the
derivative of \eq{E-k}, one obtains the group velocity of the
$\sigma$ spin mode ${v_n^\sigma }({k_x})$ that allows us to
determine a local minimum or a maximum in the energy
spectrum~\cite{Tang2012,Tang2015,Tang2017}. Solving for the spin
flip and non-flip coefficients, one uses the Landauer-B\"uttiker
formula to obtain the spin-polarized
conductance~\cite{Landauer1970,Buttiker1990}
\begin{equation}
G = g_0
  \sum_n \sum\limits_{\sigma  =  \pm }
 \sum\limits_{\sigma^\prime  =  \sigma, \bar{\sigma} }
 \frac{v_n^{\sigma^\prime}}{v_n^{\sigma}}
 \left| t_n^{\sigma \sigma^\prime } \right|^2 \, .
\label{eq3.2.30}
\end{equation}
Here $g_0=e^2/h$ is the conductance quantum per spin state of an
electron, $n$ is the number of an occupied subband, and $t_n^{\sigma
\sigma^\prime }$ indicates the transmission amplitude of the $n$th
subband electron incident from the $\sigma$ spin state scattered to
the $\sigma^\prime$ spin state.  Zero temperature is assumed.


\paragraph{Results and discussion.}

In our previous works we have presented that the finger-gate in a quantum
channel, with strong SOI in the presence of a longitudinal in-plane
magnetic field, results in bound state features in the
conductance~\cite{Tang2012,Tang2015,Tang2017}.  However, these studies
assumed that the finger gates have a delta profile and hence a mixture of
evanescent and propagating modes is not present.  Below we shall show our
numerical calculations for the investigation of top-gate bound state
features for a finite gate length $L$.

Calculations presented below are carried out under the assumption of a
2DEG at a high-mobility InAs-In$_{1-x}$Ga$_x$As interface with effective
electron  mass $m_e$ = $0.023m_0$, effective Land\'e factor $|g_s|$ =
$15$, and typical electron density $n_e$ $\sim$
$10^{12}$~cm$^{-2}$~\cite{Nitta1997}. Accordingly, the energy unit $E^*$ =
$E_\textrm{F}$ =  66~meV, the length unit $l^*$ = $1/k_\textrm{F}$ =
5.0~nm, the magnetic field unit $B^*$ = $1.14$~kT, and the Rashba
coefficient is in units of $\alpha^*$ = 330~meV~nm~\cite{Tang2012}.   We
consider channel width $W$ = $\pi l^*$ = $15.7$~nm that carries a single
propagating mode with an ideal subband bottom energy $\varepsilon_{1}$ =
$1E^*$.  Below we deal with interference of the spin-polarized electron
waves between the channel and the top-gate with negligible inter-subband
transitions based on an assumption of giant subband spacing.

\begin{figure}[t]
\begin{center}
\mbox{
 {\includegraphics[width=0.44\textwidth]{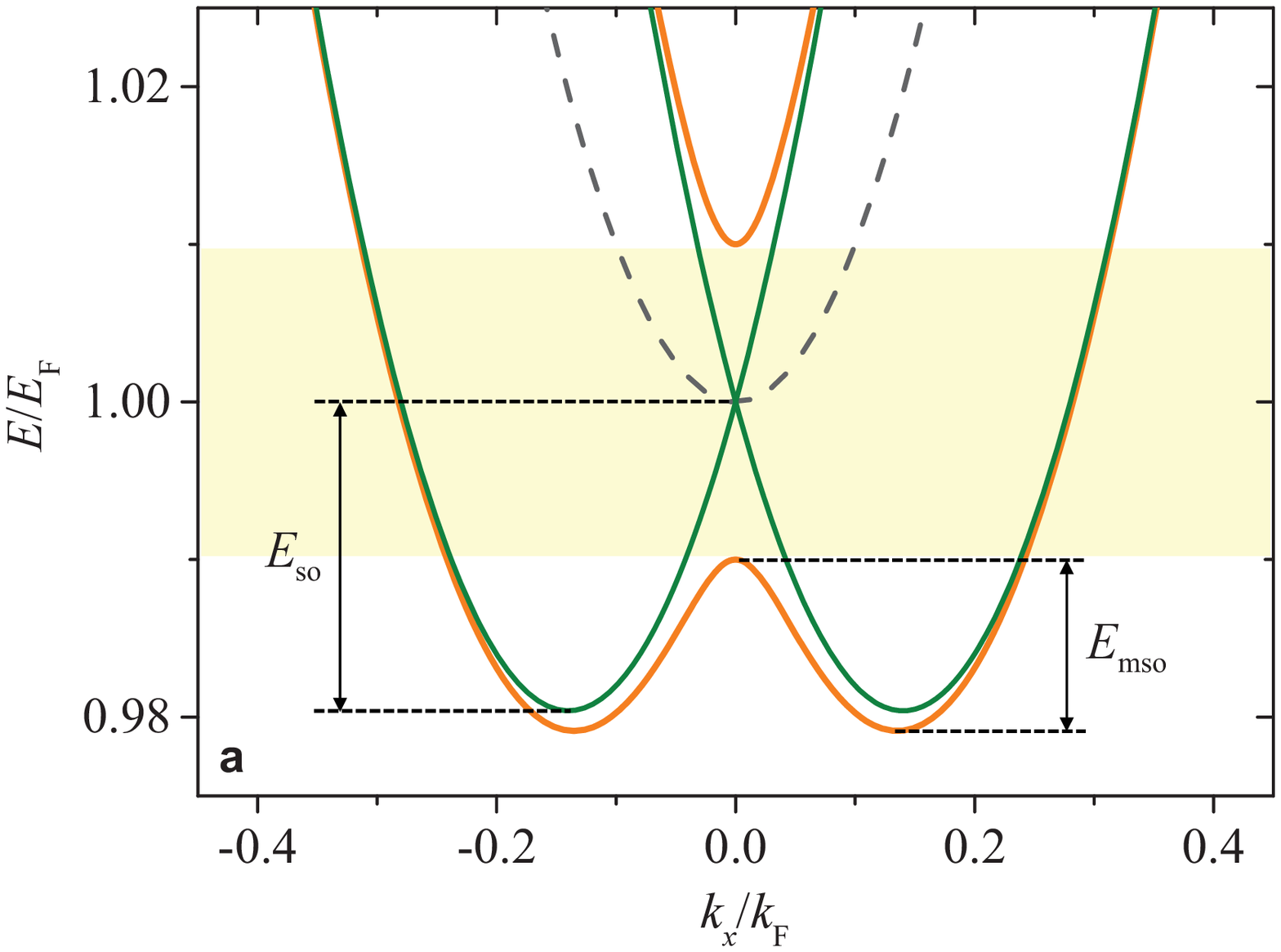}}
 }
 \mbox{
  {\includegraphics[width=0.4\textwidth]{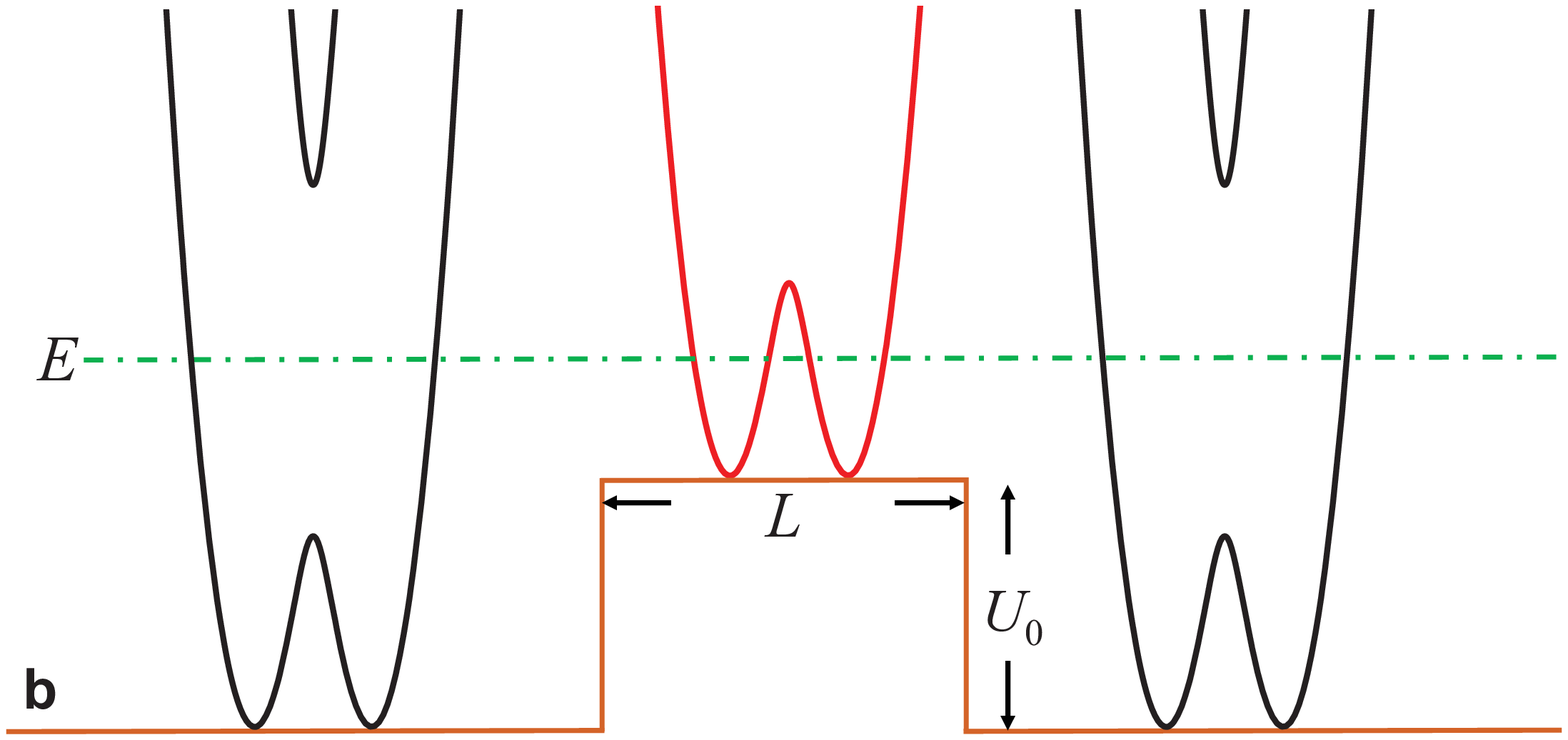}}
 }
\end{center}
\caption{Spin-split energy dispersion and the single-mode spin injection
scheme. (\textbf{a}) Dispersion relations for the lowest subband at
$\alpha$ = $gB$ = 0 (dark gray); $\alpha$ = $0.14$ at $gB$ = 0 (green)
with spin-orbit energy $E_{\rm so}$; and $\alpha$ = $0.14$ at $gB$ =
$0.01$ (orange) with magneto-spin-orbit energy $E_\mathrm{mso}$.
(\textbf{b}) Schematic illustration of the electron incident from the left
in the spin-orbit gap energy regime, through the top-gate with positive
potential energy satisfying $E_\mathrm{mso}$ $<$ $U_0$ $<$
$\Delta_\textrm{so}$, to the right lead in the gap regime.}
 \label{fig2}
\end{figure}

In \fig{fig2}(\textbf{a}) we present the lowest subband energy dispersion
obtained from \eq{E-k} for $\alpha$ = $gB$ = 0 (dash); $\alpha$ = 0.14 at
$gB$ = 0 (green); and  $\alpha$ = 0.14 at $gB$ = 0.01 (orange) with $B$ =
$1.52$~T. In the absence of a magnetic field, the spin-orbit energy
$E_{\rm so}$ = $0.0196$ corresponds to spin-orbit wave numbers
$k_\textrm{so}^\pm$ = $\mp 0.014$. The parameters satisfying the strong
SOI criterion $E_\textrm{so}$ $>$ $gB/2$ results in a spin-orbit gap
$\Delta_\textrm{so}$ = $2gB$ as shown by the shaded area.

Figure \ref{fig2}(\textbf{b}) shows the schematic of our suggested
single-mode spin injection scheme. This can be achieved using an
appropriate positive top-gate potential energy $U_0$ and the Rashba
coefficient $\alpha$.  The top-gate shifted lower spin branch in the
energy interval with magneto-spin-orbit energy $E_\textrm{mso}$ is then
located within the spin-orbit gap energy interval in the leads. We shall
show that various BS or QBS features can be observed depending on the
application of the top-gate length $L$ in this spin injection regime.

\begin{figure}[t]
\includegraphics[width = 0.45 \textwidth, angle=0] {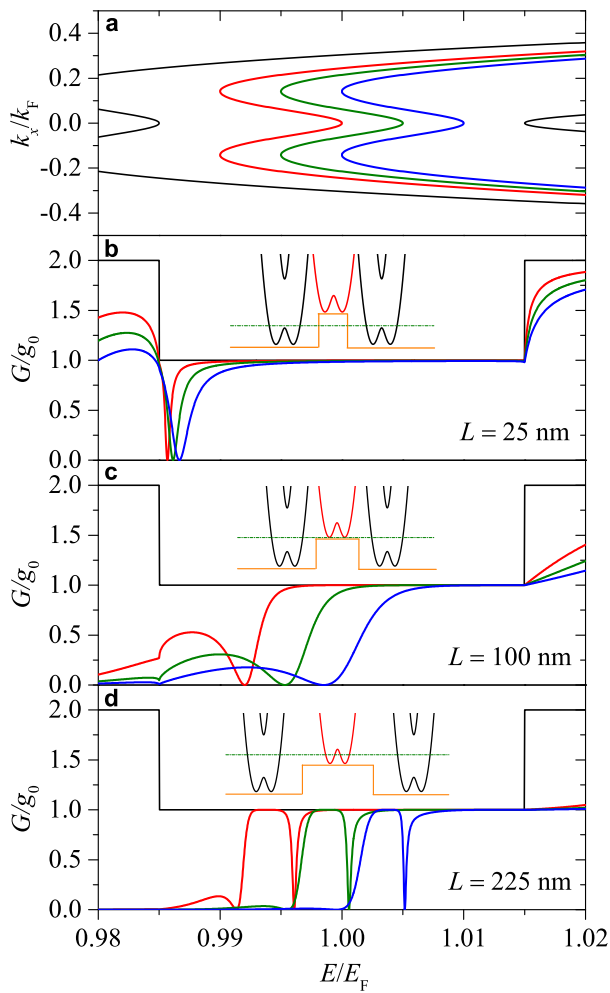}
\caption{Spinful energy dispersion and top-gate length-dependent
conductance feature. (\textbf{a}) Dispersion relation of the lowest
subband for $U_0$ = 0 (black), 0.015 (red), 0.020 (green), and 0.025
(blue). Corresponding energy characteristics of conductance are presented
for gate-length (\textbf{b}) $L$ = 25 nm, (\textbf{c}) $L$ = 100 nm, and
(\textbf{d}) $L$ = 225 nm. The Rashba coefficient $\alpha$ = $0.15$ at
$gB$ = $0.015$. Inset: various bound-state features.}
 \label{fig3}
\end{figure}

For illustration \fig{fig3} shows (\textbf{a}) the lowest subband energy
dispersion and (\textbf{b})-(\textbf{d}) conductance spectra of various
top-gate potential energies $U_0$ = 0.015 (red), 0.020 (green), and 0.025
(blue) for the Rashba coefficient $\alpha$ = $0.15$ at $gB$ = $0.015$.
Here the magneto-spin-orbit energy $E_\textrm{mso}$ = 0.01 and the
spin-orbit gap energy $\Delta_\textrm{so}$ = 0.03 in order to inject
electrons from the spin-orbit gap to the low-energy regime, satisfying the
criterion $E_\textrm{mso}$ $<$ $U_0$ $<$ $\Delta_\textrm{so}$ where the
strength of top-gate potential is between the magneto-spin-orbit energy
and the spin-orbit gap.

For the application of top-gate with short length, the electron modes
occupying the subbands in the leads dominate the transport properties. For
applying a positive top-gate potential energy, the conductance spectrum
thus reveals a hole-like QBS feature located around the lower spin branch
top $E_\textrm{top}^-$ = 0.985 in the leads, as depicted in
\fig{fig3}(\textbf{b}) and in the inset.  This QBS feature can also be
found finger-gate systems~\cite{Tang2012} or refers to the Fano-Rashba
resonances~\cite{David06,David08}. Our results imply that the finger-gate
model is valid if the length of the top-gate is shorter than the Fermi
wave number.

For an intermediate top-gate length ($L$ = 100 nm), the hole-like QBS
feature shown in \fig{fig3}(\textbf{b}) is suppressed and reduces to a
knickpoint. Instead, the electron occupying the outer propagating mode in
the spin-orbit gap in the leads may be scattered into the lower spin
branch bottoms in the top-gate (see \fig{fig3}(\textbf{c}), inset). As a
result, the electron is allowed to make spin-flip transitions forming an
electron-like QBS in the top-gate around the energy
\begin{equation}
E_\textrm{bottom}^-(gB,U_0) = {\varepsilon_1} - {\alpha ^2}-
\left(\frac{gB}{2\alpha}\right)^2 + U_0
\end{equation}
that is related to the Zeeman factor $gB$ and shifted by the top-gate
potential energy $U_0$.  This transport signature exhibits the coupling of
an outer propagating mode in the leads and evanescent modes around the
bottoms of the $U_0$ shifted lower spin branch in the top-gate. The
conductance valley broadens with increasing $U_0$. Suppose the length of
the top-gate is between short and intermediate situations, such as $L$ =
60 nm (not shown), the conductance drops become broader and locate between
the lead band top and the top-gate band bottom of the lower spin branch.

Now we turn to consider the top-gate with a sufficient long length ($L$ =
225 nm) as shown in \fig{fig3}(\textbf{d}).  We see that the electron-like
QBS feature shown in \fig{fig3}(\textbf{c}) is strongly suppressed.  A
sufficient long $L$ significantly enhances the electron dwell time in the
top-gate region and favors multiple scattering. As a result, a hole-like
BS feature can be observed if the electron is injected into the top-gate
region with energy at the reverse point of the inner mode in the top-gate
shifted lower spin branch.  Figure \ref{fig3}(\textbf{d}) shows clear
sharp dip structures in the conductance spectra with conductance zeros.
Here the electrons, occupying the inner modes below the $U_0$ shifted
subband top energies $E_\textrm{top}^-(U_0)$ = $0.985$ + $U_0$, behave a
hole-like BS feature at $E/E_\textrm{F}$ = 0.996, 1.001, and 1.005 for
$U_0$ = 0.015, 0.020, and 0.025, respectively. These hole-like BS energies
are at the reverse point energies
\begin{equation}
E_{{\rm{rev}}}^- = {\varepsilon _1} + {U_0} - {\left(
{\frac{{gB}}{{2\alpha }}} \right)^2} + {\left[ {\frac{{{{\left( {gB}
\right)}^2}}}{{4\alpha }}} \right]^{2/3}} - {\left[ {2{\alpha
^2}{{\left( {gB} \right)}^2}} \right]^{1/3}}
\end{equation}
of the inner modes in the top-gate shifted lower spin branch, as depicted
in the inset. Here the inverse of the second derivative of the energy
dispersion is divergent leading to a divergent effective mass.


\paragraph{Concluding remarks.}

We have considered a top-gate controlled quantum device with spin-orbit
coupling and an external in-plane magnetic field. Since the effective
spin-orbit field $B_\textrm{so}$ and Zeeman field $B$ are perpendicular,
the spin-orbit gap can be induced in the strong spin-orbit coupling
regime. Transport signatures have been demonstrated for electrons incident
from spin-orbit gap energies though a top-gate with positive electric
potential energies. Our theoretical calculations suggest possible
conditions and mechanisms of an electron-like QBS, a hole-like QBS, or a
hole-like BS feature in continuum. Our calculations can be used to extract
information of spin-orbit gap involved top-gate tunneling spectroscopy
experiments, such as the recently reported measurements in Heedt
\textit{et al}.~\cite{Heedt17}.


 \begin{acknowledgments}
The authors acknowledge financial supports from Ministry of Science
and Technology in Taiwan under grant No.\ 106-2112-M-239-001-MY3,
the Research Fund of the University of Iceland, the Icelandic
Research Fund under grant No.\ 163082-051, and the Icelandic
Instruments Fund.
 \end{acknowledgments}


\end{document}